\documentclass[conference]{IEEEtran}
\IEEEoverridecommandlockouts
\usepackage{array,multirow}
\newcolumntype{P}[1]{>{\centering\arraybackslash}p{#1}}
\renewcommand{\baselinestretch}{0.98}
\usepackage{cite}
\usepackage{amsmath,amssymb,amsfonts}
\usepackage{subcaption}
\usepackage{graphicx}
\usepackage{textcomp}
\usepackage{xcolor}
\def\BibTeX{{\rm B\kern-.05em{\sc i\kern-.025em b}\kern-.08em
    T\kern-.1667em\lower.7ex\hbox{E}\kern-.125emX}}
\begin{document}

\title{Social Media Brand Engagement as a Proxy for E-commerce Activities: A Case Study of Sina Weibo and JD\\
}

\author{\IEEEauthorblockN{Weiqiang Lin\IEEEauthorrefmark{1}, Pedro Saleiro\IEEEauthorrefmark{2}, Natasa Milic-Frayling\IEEEauthorrefmark{3}, Eugene Ch'ng\IEEEauthorrefmark{4}}
\textit{\IEEEauthorrefmark{1}International Doctoral Innovation Center, University of Nottingham, Ningbo, China}\\
\textit{\IEEEauthorrefmark{2}Department of Computer Science, University of Chicago, Chicago, USA} \\
\textit{\IEEEauthorrefmark{3}School of Computer Science, University of Nottingham, Nottingham, UK}\\
\textit{\IEEEauthorrefmark{4}NVIDIA Joint-Lab on Mixed Reality, University of Nottingham, Ningbo, China}\\
\{wyatt.lin, eugene.chng\}@nottingham.edu.cn, Natasa.Milic-Frayling@nottingham.ac.uk, saleiro@uchicago.edu\\
}

\maketitle

\renewcommand{\baselinestretch}{1}
\newcommand\todo[1]{\textcolor{red}{#1}}

\title{Social Media Brand Engagement as a Proxy for E-commerce Activities: A Case Study of Sina Weibo and JD.com}
\begin{abstract}

E-commerce platforms facilitate sales of products while product vendors engage in Social Media Activities (SMA) to drive E-commerce Platform Activities (EPA) of consumers, enticing them to search, browse and buy products. The frequency and timing of SMA are expected to affect levels of EPA, increasing the number of brand related queries, clickthrough, and purchase orders. This paper applies cross-sectional data analysis to explore such beliefs and demonstrates weak-to-moderate correlations between daily  SMA and EPA volumes. Further correlation analysis, using 30-day rolling windows, shows a high variability in correlation of SMA-EPA pairs and calls into question the predictive potential of SMA in relation to EPA. Considering the moderate correlation of selected SMA and EPA pairs (e.g., Post-Orders), we investigate whether SMA features can predict changes in the EPA levels, instead of precise EPA daily volumes. We define such levels in terms of EPA distribution quantiles (2, 3, and 5 levels) over training data. We formulate the EPA quantile predictions as a multi-class categorization problem.  The experiments with Random Forest and Logistic Regression show a varied success, performing better than random for the top quantiles of purchase orders and for the lowest quantile of search and clickthrough activities. Similar results are obtained when predicting multi-day cumulative EPA levels (1, 3, and 7 days). Our results have considerable practical implications but, most importantly, urge the common beliefs to be re-examined, seeking a stronger evidence of SMA effects on EPA.
\end{abstract}




\maketitle

\section{Introduction} \label{sec:intro}

E-commerce platforms enable promotion and sales of products at large scales and resort to careful monitoring and optimization of product sales cycle in order to ensure quality services to both sellers and buyers. Thus, a substantial effort has been put into studying clickstreams and predicting product purchases \cite{yeo2017predicting,lo2016understanding,moe2003buying,sismeiro2004modeling,park2015predicting}. Clickstream data typically includes searching and browsing for specific brands or products and clicks on product information for feature and price comparison. Patterns of such user activities have been successfully used to predict users' intent to purchase once engaged with the platform \cite{moe2003buying, van2005predicting, bertsimas2003dynamic, park2015predicting, sismeiro2004modeling}. Indeed, searching, browsing, and comparing products are aligned with the Purchase Decision Models (PDM) used in marketing and sales management \cite{kotler2015book} and therefore, good predictors of consumer purchases. However, vendors are equally interested in creating awareness of their brands and products to drive traffic to their Web sites and e-commerce platforms. 
\begin{figure}[h]
    \captionsetup{justification=centering}
    \centering
    \textbf{}\par\medskip
	\includegraphics[width=\linewidth]{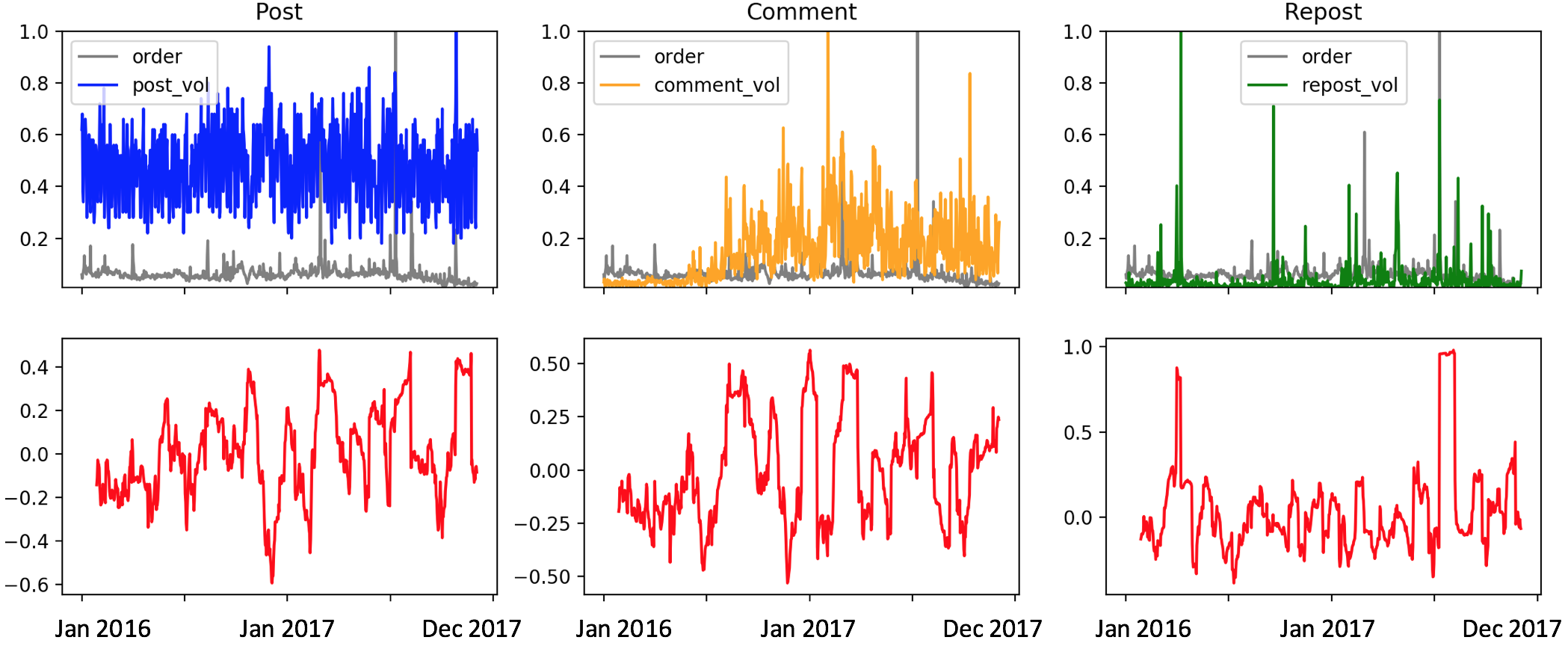}
    \caption{Rolling correlation between SMA on day $t_p$ and EPA on day $t_{p+1}$, with a 30-day rolling window over two years.}\label{fig:huawei_post}
\end{figure}
This has led to increased commitment to Social Media Monitoring (SMM) by vendors and, consequently, to the development of tools and services for marketeers to manage their Social Media Activities (SMA). The increased investment in SMM and SMA is driven by an underlying belief that social media is a key factor in business success, particularly for product promotion and sales. However, considering the global adoption of e-commerce platforms, it is important to examine that assumption and investigate to what degree vendors' SMA affects consumer engagement with their products, i.e., the consumers' E-commerce Platform Activities (EPA). In fact, we expect that the effects of vendors' SMA are conflated with and dominated by advertising campaigns organized by the e-commerce platform itself and global retail events, such as Black Friday, 11-11, and June 18 in China. At the same time, EPA analyses and predictions of SMA effects are out of reach of individual vendors and marketeers as they require comprehensive data collections and analysis and sophisticated machine learning methods. 

Our research involves a cross-sectional data analysis of multiple time-series from SMA and EPA and a new approach to predicting EPAs, formulated as a multi-class categorization problem with classes defined by discretized time-series spectra. More precisely, we specify distinct levels of EPA activities that correspond to quantiles of the EPA data distribution over a given time period. We use a data set provided by JD e-commerce platform (JD.com) that comprises search, clickthrough and product orders for a sample of 33 vendors from five product categories. For each vendor in the sample we collect social media posts, reposts, and commentaries published on the Sina Weibo (Weibo.com) social media platform. The data covers a period of two years. We use time-alignment of vendors' social media posts, reposts, and comments with EPA that correspond to different decision stages in the consumers' purchases: search, clickthrough, and orders. By relating the vendors' SMA and EPA we postulate: 

\textit{Should the data analysis reveal high correlation between a vendor's daily SMA and its customers' daily EPA, we might be able to create reliable predictors of EPA based on SMA features.} 

Our correlation analyses of specific SMA and EPA pairs, i.e., Pearson correlation of Post with Search, Clickthrough with Order, and Repost with Comment data for different lag factors (1-15 day shifts), point to a highly variable correlation over time, as illustrated in Figure~\ref{fig:huawei_post}. Across the sample of 33 vendors, the correlation coefficients of SME-EPA pairs vary between 0.08 and 0.56, indicating that predicting daily volumes of EPA purely based on SMA would not be reliable. Nevertheless, the question still remains:

\textit{Could SMA features be used to predict changes in the levels of EPA, i.e., highs and lows in search, clickthrough and orders, especially if SMA has a sustained effect over several days, e.g., 3 to 7 days following a specific social media activity?}

In order to explore this, we create predictive models for specific levels of cumulative search, clickthrough, and orders added up over 1, 3, and 7 days, where levels are defined by the quantiles of the corresponding distributions over the training time period. We apply Random Forest and Logistic Regression classifiers for categories corresponding to 2, 3, and 5 quantiles, i.e., (i) above the median, (ii) within three quantiles: (0-33\%, 33-66\%, 66-100\%), and (iii) within five quantiles (0-20\%, 20-40\%, 40-60\%, 60-80\%, and 80-100\%). 

Our experiments show that SMA can be used to predict top quantiles of product orders and low quantiles of search and clickthrough across vendors' categories. The performance for the remaining ranges is on par with random predictions. The results are important for optimizing the e-commerce platform operations for peaks in EPA and for evaluating marketing campaigns through SMA. From the research perspective, our work is the first to provide a detailed analysis of SMA-EPA pairs and a method that reveals selective roles that SMA types play in predicting levels of specific EPA types.

\section{Related work}

Most of the existing work in understanding purchase behavior and user intent aims at estimating conversion rates for individual products, individual customers, or both \cite{xu2014path,  moe2003buying, van2005predicting, bertsimas2003dynamic, park2015predicting, sismeiro2004modeling, yeo2017predicting}. Computational studies of consumers behavior have adopted ad-hoc approaches and brute-force feature engineering, aiming to uncover factors that explain user behavior, particularly the intent to buy {\cite{lo2016understanding, park2015predicting,yeo2017predicting}}. In contrast, studies within business and market research have led to conceptual frameworks that are useful for both optimizing and interpreting experiment results, particularly the effects of features used in predictive models
{\cite{van2005predicting,sismeiro2004modeling}}.

Considering e-commerce platforms, research has focused on  behavioral patterns, product characteristics \cite{moe2003buying,bertsimas2003dynamic} and feedback in product reviews \cite{yu2012mining}, and use them as features for predicting the online product sale. Such predictive features often originate from resources internal to the shopping platform \cite{young2004predicting, moe2003buying}, or those collected externally such as social media content \cite{hawkins2012consumer} or query logs from online search engines \cite{kulkarni2012using}. 

With regards to social media, past studies explored how commercial intent, expressed in tweets, correlates with external events \cite{wang2013mining,hollerit2013towards}. Homophily was found to have a significant influence on the purchase intent, i.e., users are more likely to purchase products similar to those purchased by their friends due to word-of-mouth effects \cite{kooti2016portrait}. This phenomenon is important for e-commerce and has been studied in detail over the past decade \cite{leskovec2007dynamics,anagnostopoulos2008influence}. It has also been recognized that other external factors, such as geographic proximity, may explain similarity in purchase patterns among friends on social medial \cite{crandall2008feedback,shalizi2011homophily}. Furthermore, text analysis has been applied to detect purchase intent from social media content \cite{gupta2014identifying} and Facebook profiles \cite{zhang2013predicting} and provide recommendations based on micro-blogging activities \cite{zhao2014we}. 

We complement this work by considering social media monitoring practices and macro-level analyses that do not involve individual user profiles and content analysis but focus on the levels and timing of brand-specific social engagements, i.e., volumes of vendors' post and the followers commentaries and repost. Understanding temporal patterns of user activities has generated a wide interest over the past decade, including the identification of broad classes of temporal patterns based on activity peaks \cite{crane2008robust,lehmann2012dynamical, romero2011differences, yang2011patterns}. Usually, those classes are defined based on specific volumes ranges and duration of activities before and after the peak. Crane and Sornette \cite{crane2008robust} define endogenous and exogenous origins of peaks based on triggers generated by internal aspects of the social network. They found that popularity is mostly driven by exogenous factors instead of endemic spreading. Yang and Leskovec \cite{yang2011patterns} propose a new measure of time series similarity and clustering. 

In our research we apply cross-sectional analysis of data from two distinct platforms rather than studying specific aspects of commercial intent within social media or e-commerce platforms themselves. With the aim to explore the two sets of temporal data, we apply correlation analysis and formulate discretized prediction problems that can leverage SMA features to predict certain levels of e-commerce activities despite a highly volatile correlation of the corresponding data series. 
\section{Preliminaries and Data Analysis}   \label{sec:prel}

E-commerce platforms are instrumented to capture information about shopping activities and gather detailed statistics about consumers' interactions with facilities that support the purchase workflow. Analyzing such data is instrumental for optimizing the platform operations. Our research is conducted in collaboration with JD\footnote{JD.com is one of the largest e-commerce platforms in China.} with access to anonymized data about product sales of vendors in different product categories. Furthermore, we include information about vendors' social engagements on Sina Weibo\footnote{Weibo.com is the largest social media platform in China, with social media features that combine those of Facebook (www.facebook.com) and Twitter (www.twitter.com).} platform that hosts Web sites of many JD vendors. The vendors share post and engage the community through comment and discussions.  We align Sina Weibo data and JD logs to understand how the vendors' Social Media Activities (SMA) on Weibo relate to the E-commerce Platform Activities (EPA) of JD users, focusing on patterns of interaction within both services. We perform a macro-level analysis considering time-series of actions; analyses of individual users' behavior or content shared in social media are not part of this study. 

\begin{table}[h]
\captionsetup{justification=centering}
\caption{Total social media activities of the sampled vendors within each product category for the two year period (Jan 2016-Dec 2017).}\label{Tab:datasetstatistics}
\centering
\begin{tabular}{l|P{2.3cm}|c|c|c}
\hline
Category & Number of Vendors& Post& Comment& Repost\\ 
\hline\hline
Phone\&El&4&37,312&5,143,200&2,629,674\\
Sports&5&6,927&295,952&230,733\\
Food&9&17,666&1,072,441&496,664\\
Clothes&6&9,456&227,258&155,334\\
Home &9&53,857&1,902,164&681,071\\
\hline
\end{tabular}
\end{table}

\begin{figure}[h]
    \captionsetup{justification=centering}
    \centering
    \textbf{}\par\medskip
	\includegraphics[width=\linewidth]{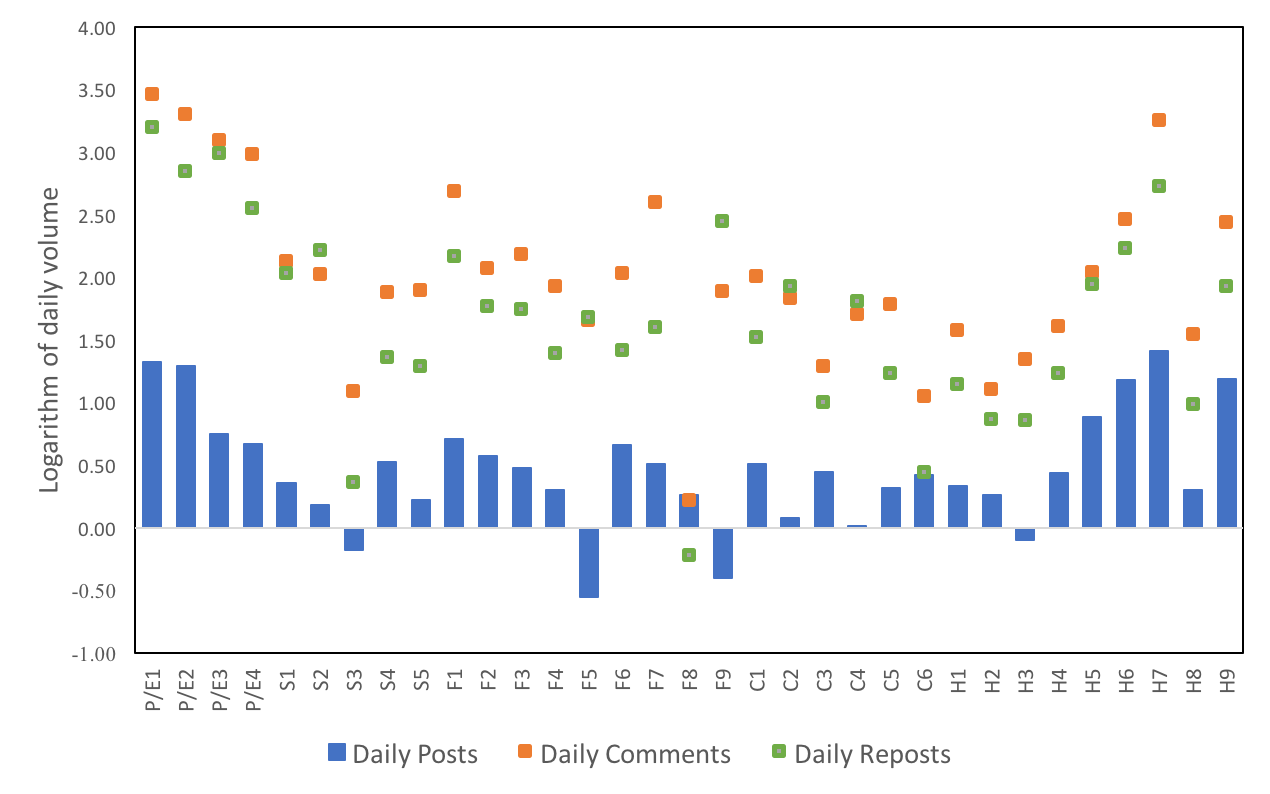}
    \caption{Total Post, Comment and Repost statistics normalized by the corresponding maxima over the two year period (Jan 2016-Dec 2017).}\label{fig:daily_sma_line}
          \vspace{-1.5em}
\end{figure}

\begin{figure}[h]
    \captionsetup{justification=centering}
    \centering
    \textbf{}\par\medskip
	\includegraphics[width=\linewidth]{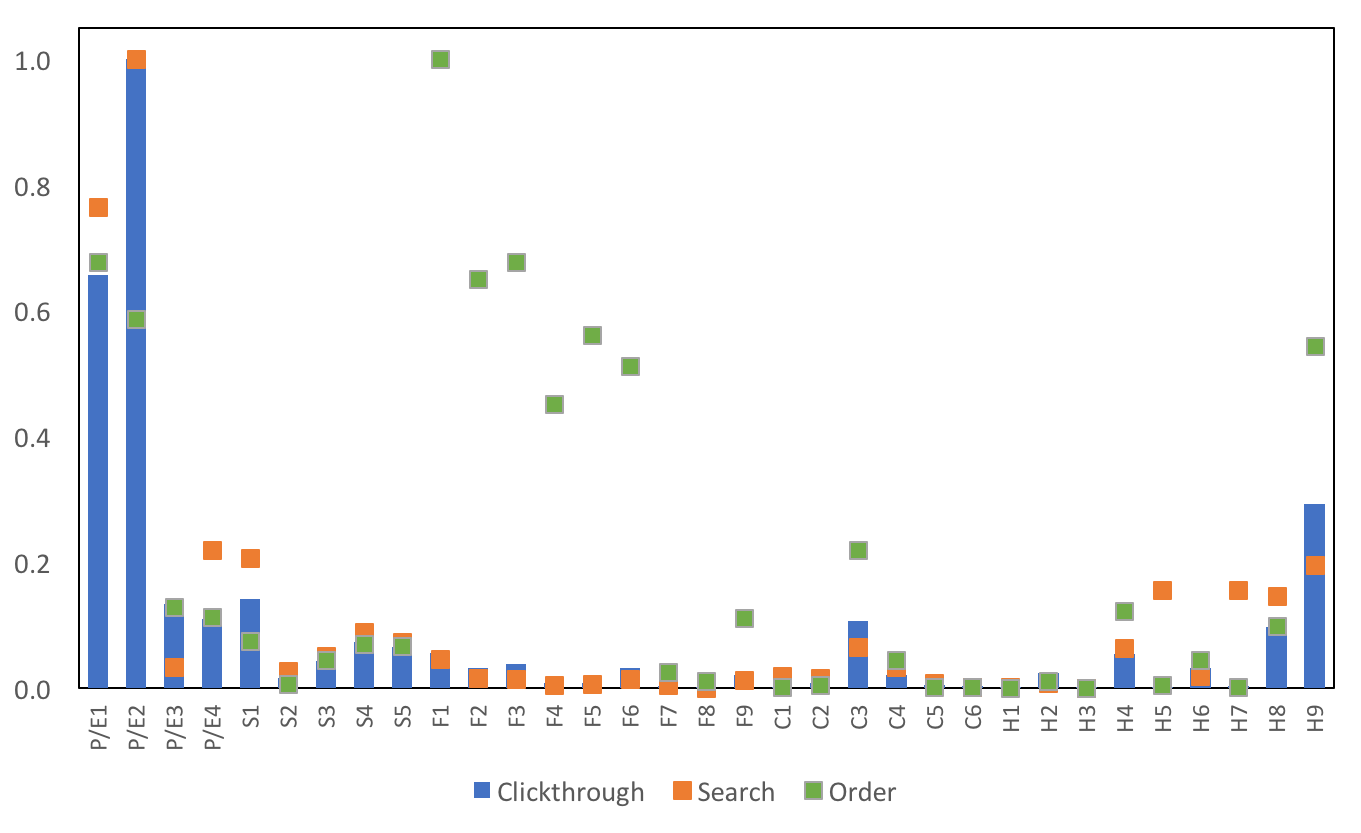}
    \caption{Total Search, Clickthrough and Order volumes of individual vendors normalized by the daily maxima over the two year period (Jan 2016-Dec 2017).}\label{fig:epa_stats}
          \vspace{-1.5em}
\end{figure}

\subsection{Data}   \label{ssec:data}
Our EPA data set comprises JD records of (1) Search for brands, (2) Clickthrough, and (3) Orders. Search queries are included only if explicitly mention a vendor's name. The data was collected for a sample of 50 \textit{most popular} JD vendors across 5 product categories: Phone\&Electronics, Sports, Food, Clothes and Home on JD's platform. The vendor's popularity ranking is based on the JD's sales performance metrics that is treated as business confidential. Among 50 companies we identified 33 that have had a Weibo account for at least two years, from Jan 2016 until Dec 2017, and published at least 10 posts over that period. The final sample comprises: 4 vendors in Phone\&Electronics (P/E), 5 in Sports, 9 in Foods,  6 in Clothes, and 9 in Home. 
For each vendor we collected: (1) Post, (2) Repost, and (3) Comment statistics.

The EPA and SMA data is segmented per calendar date. Table~\ref{Tab:datasetstatistics} shows the SMA distribution per vendor category and Figure~\ref{fig:daily_sma_line} plots the total SMA volumes of Post, Repost and Comment activities for individual vendors on a logarithmic scale. As expected, the Repost and Comment statistics are two orders of magnitude  higher than Post, illustrating a strong social media effect in response to the vendors' posting activities.  Considering the SMA statistics across five categories (Figure~\ref{fig:daily_sma_line}), the vendors in the Phone\&Electronics category appear to gain most traction through repost and commenting activities. Sports category ranks lowest on posts while the Clothes category is the lowest when considering the sum of all three SMA statistics. 

Figure~\ref{fig:epa_stats} presents the total Search, Comment and Order statistics for individual vendors, normalized by the maximum daily volume over the period of two years. It shows that online purchases incur Search and Clickthrough volumes with 1-2 orders of magnitude higher than Order volumes as they play important parts in the purchase decision process. 

\begin{figure*}[h]
    \captionsetup{justification=centering}
    \centering
    \begin{subfigure}[t]{0.32\textwidth}
        \centering
        \includegraphics[width=\linewidth]{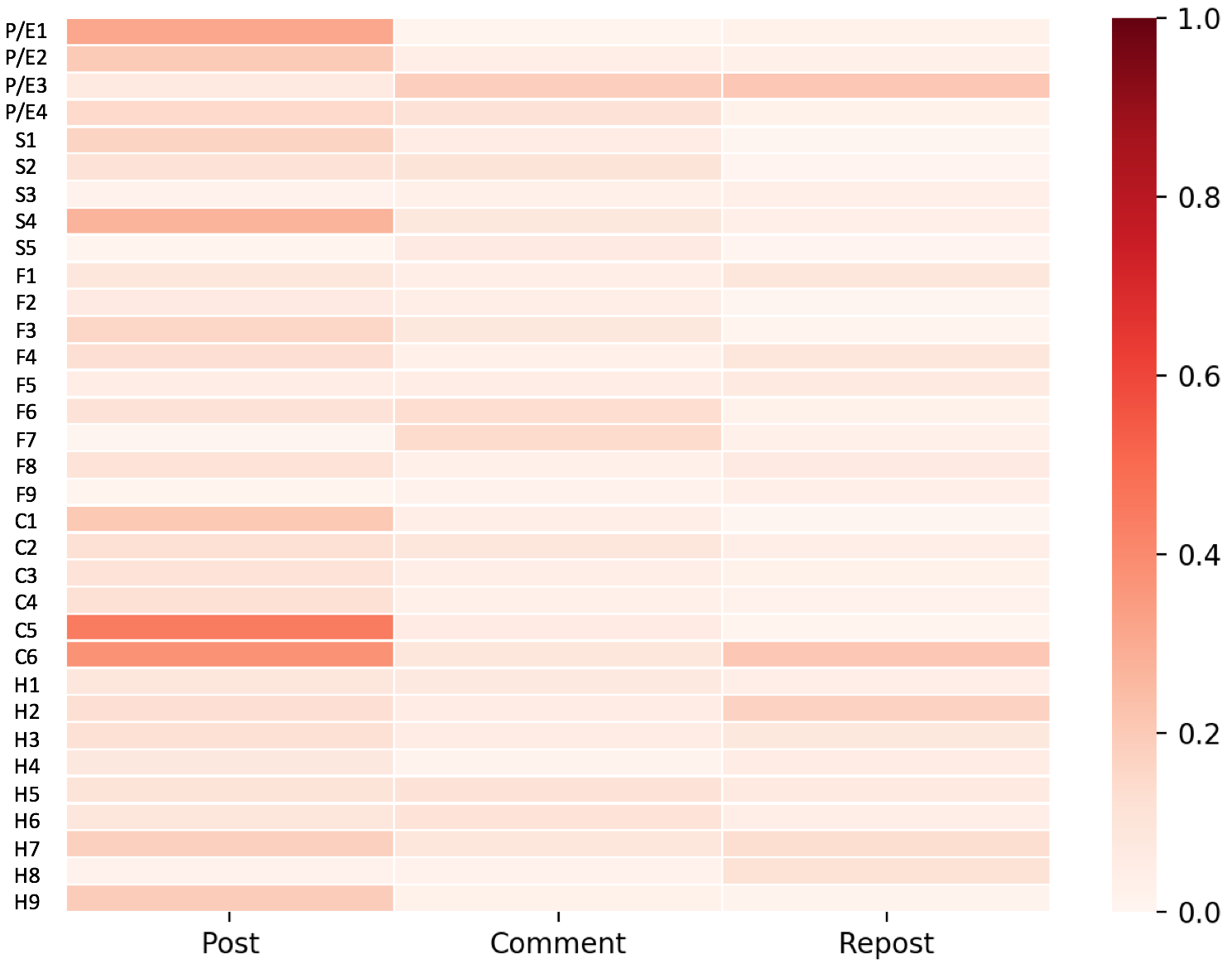} 
        \caption{Pearson correlations between SMA on day $t_p$ and Search on $t_{p+1}$} \label{fig:search2sma}
    \end{subfigure}
    \hfill
    \begin{subfigure}[t]{0.32\textwidth}
        \centering
        \includegraphics[width=\linewidth]{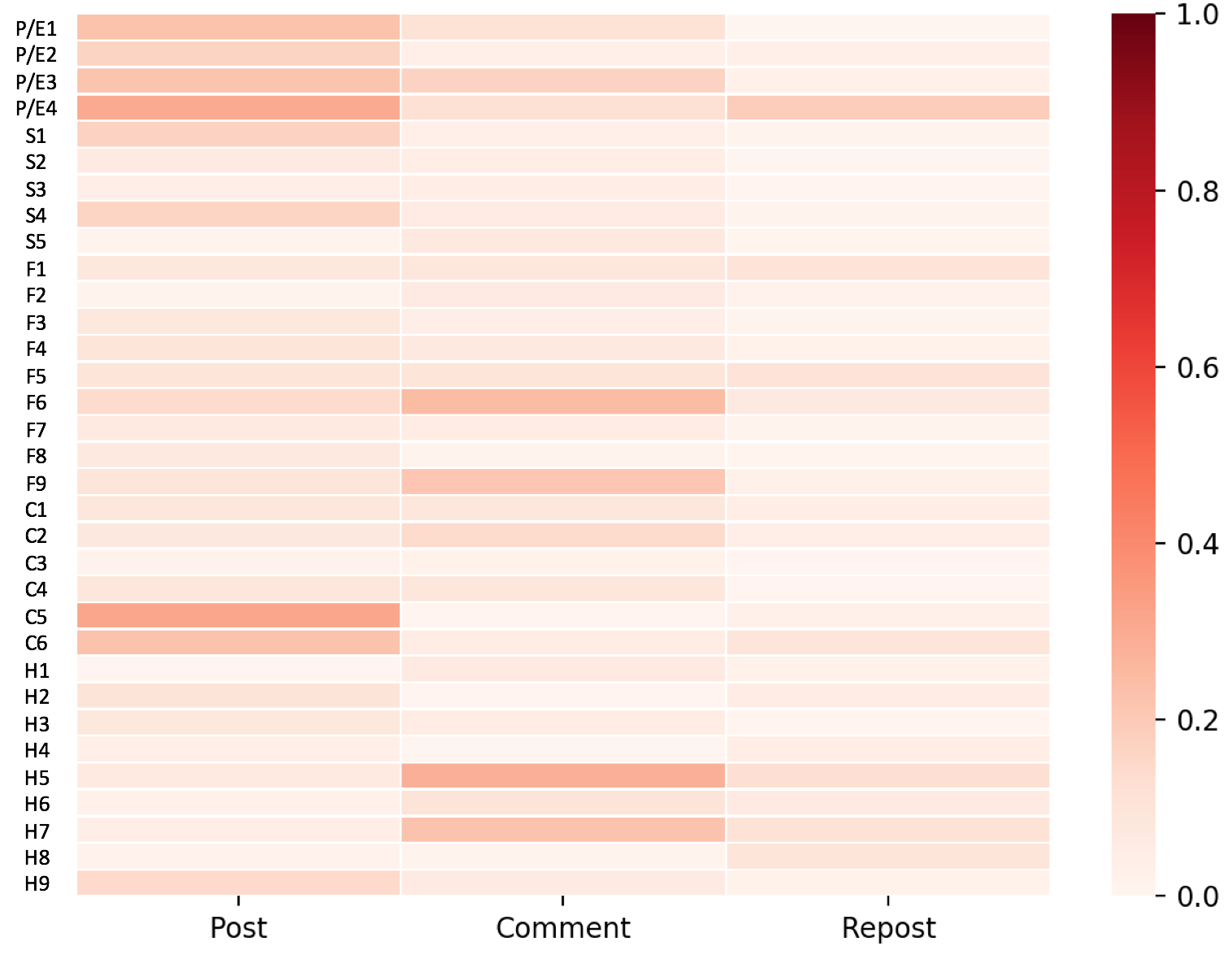} 
        \caption{Pearson correlations between SMA on day $t_p$ and Clickthrough on $t_{p+1}$} \label{fig:clickthrough2sma}
    \end{subfigure}    
    \hfill
\begin{subfigure}[t]{0.32\textwidth}
        \centering
        \includegraphics[width=\linewidth]{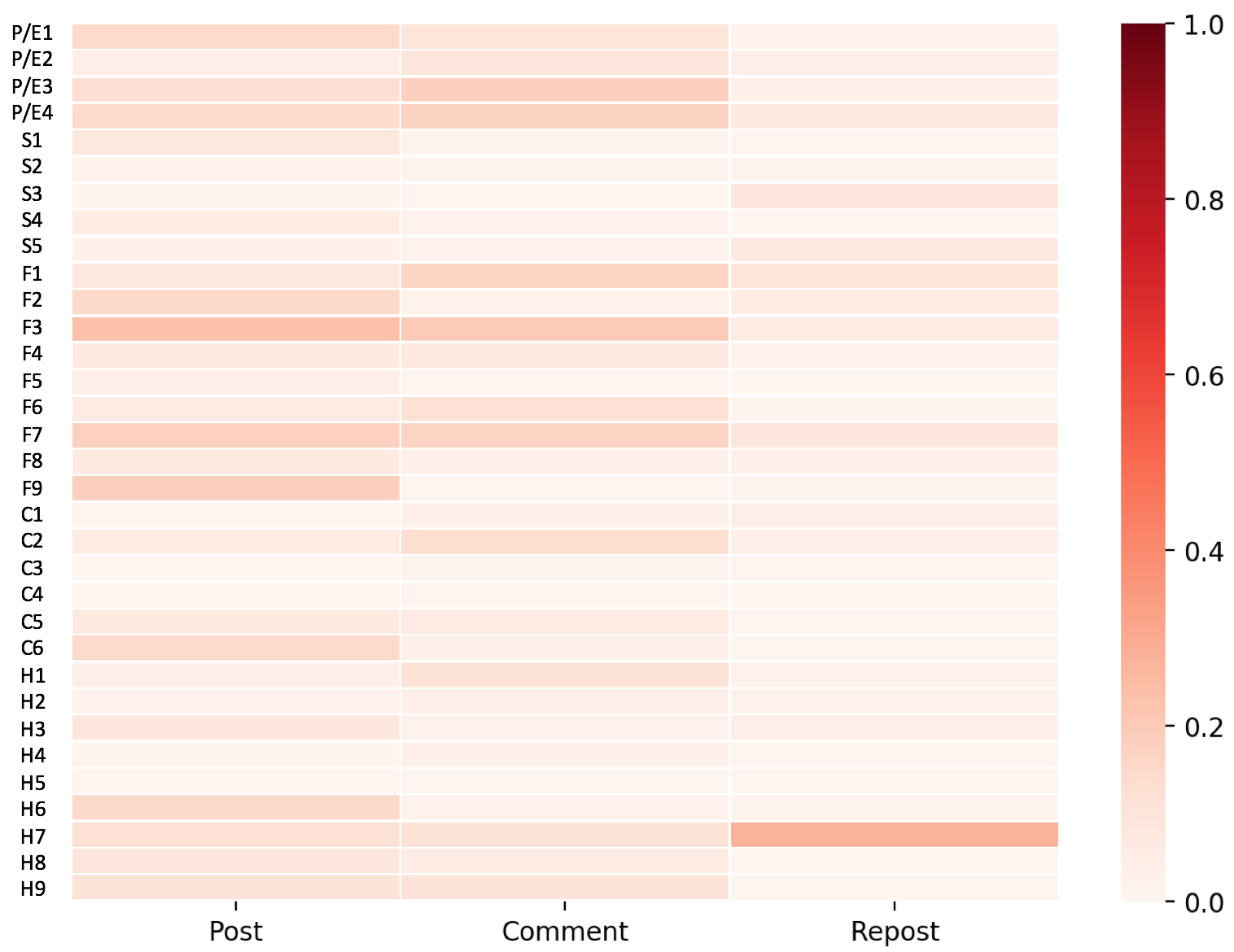} 
        \caption{Pearson correlations between SMA on day $t_p$ and Order on $t_{p+1}$} \label{fig:ord2sma}
    \end{subfigure}
    \caption{Pearson Correlation between social media activities (Post, Repost, and Comment) and specific e-commerce platform activities (Search, Clickthrough and Order) for 33 vendors using their two year time series. Heat maps show low to medium correlations between SMA and EPA pairs.}\label{fig:epa2sma}
\end{figure*}

\subsection{Correlation Analysis}

We calculate the pairwise Pearson Correlation between SMA (Post, Repost, Comment) and EPA (Search, Clickthrough, Order) for the two year time-series  corresponding to the individual vendors. Table~\ref{Tab:highest_corr}  shows the highest correlation coefficient for each SMA-EPA pair among all 33 vendors. It confirms low-to-medium correlations of SMA-EPA across vendors, with Post-Search having the highest maximum coefficient of 0.56. The maximum of 0.11 for Repost-Clickthrough and 0.08 for Repost-Orders indicate low correlations between these SMA and EPA types across all the vendors.

\begin{table}[h]
\captionsetup{justification=centering}
\caption{The highest correlation coefficients across SMA-EPA pairs for the sample of 33 vendors.\iffalse \textcolor{red}{Transpose the table. Post, Repost and Comment on the side and Search, Clickthrough and Orders on the top. Keep this order precisely }\fi}\label{Tab:highest_corr}
\centering
\begin{tabular}{l|P{1.8cm}|P{1.8cm}|P{1.8cm}}
\hline
 &  Post& Comment& Repost\\ 
\hline\hline
Clickthrough&0.39&0.23&0.11\\
Search&0.56&0.20&0.23\\
Order&0.25&0.21&0.08\\
\hline
\end{tabular}
\end{table}

Heat maps in Figure~\ref{fig:epa2sma} present the pairwise SMA-EPA correlation coefficients for each of the 33 vendors across the full two year period by considering the next-day EPA statistics for a given SMA day statistics. Comparing the relative importance of SMA types, it appears that the Post statistics are more highly correlated with Search and Clickthrough, driving the brand awareness and product interest. Considering Comment statistics across vendors, they are more highly correlated with Clickthrough and Orders than  with Search, suggesting that comments on posts and reposts may help with purchase decisions. Our analysis is the first to offer empirical evidence that different types of social media engagements relate to different aspects of online shopping activities. 

We further consider SMA-EPA correlations for different time frames and time lags.


\subsubsection{Rolling Correlation Analysis}

We calculate the Pearson Correlation Coefficient for SME-EPA pairs using the rolling 30-day window over the period of two years (instead of the single two-year time-series) and show that the correlation coefficients for daily statistics varies over time. This is illustrated in Figure~\ref{fig:huawei_post} (Section~\ref{sec:intro}), showing time variations of the correlation coefficients for the vendor with the highest volume of comments. Without stable correlations it would be hard, for example, to use Comment features to predict Orders.     


\subsubsection{SMA-EPA Lag Analysis}

In our scenario, SMA and EPA occur on different platforms, i.e., Sina Weibo and JD, respectively. Thus, even if SMA has an effect on EPA, one can expect delays, depending on the speed of social media propagation. For that reason, we investigate rolling SMA-EPA correlations with different day lags, allowing for 1-15 day delay. By repeating the 30-day rolling window calculations with 1-15 day shift in time series, we observe a weekly pattern in the rolling correlation, with a spike every 7 days. However, even for the vendor with the highest volume of Comment activities, the absolute correlation values are always below 0.5. Generally, the volumes of Post activities have the highest correlation with the volumes of Search or Clickthrough, while Comment data has the highest correlation with Order.

\subsection{Problem Formulation}
Concluding from the presented analysis, the task of predicting daily volumes of e-commerce activities from levels of social media activities is not well supported by the correlation analysis of daily statistics. However, in many practical scenarios it is sufficient to predict a trend, i.e., a change in the volume range, anticipating highs and lows that may affect product supplies and platform logistics. Furthermore, instead of daily activities, it is sufficient to predict purchase outcomes over a given period of time, e.g., a cumulative sales for 3-7 ahead. Thus, in Section~\ref{sec:formulation} we formulate the EPA prediction problem by considering 
(a) discretized distributions of EPA volumes for each vendor into 2, 3, and 5 quantiles and (b) quantile predictions of EPA totals for the next day, next 3 days and next 7 days.

\section{Predicting E-commerce Activities} \label{sec:formulation}

Let us now consider a set of vendors, represented by their brands $B = \{b_1, b_2, ..., b_i, ...\}$. Without a loss of generality, we can assume that each vendor is represented by a single brand that has a specific daily stream of social media signals $S_t = \{s_1, s_2, ..., s_j, ...\}$, consisting of official Post activities by the vendor and the Repost and Comment activities by the users of the social media platform. Similarly, each brand has a daily stream of e-commerce platform activities (EPA) $E_t = \{e_1, e_2, ..., e_l, ...\}$ corresponding to Search, Clickthrough and Order actions on a given day $t$. Some of the social media users may also be customers of the e-commerce platform but we do not attempt to align the user activities across the two platforms. 

We consider the predictive power of the social media signals regarding specific e-commerce platform activities, i.e., $f_v(b_i, t, S_t, E_t)$, a discrete volume function of a brand $b_i$, a timestamp $t$, a social media stream $S_t$ and a type of e-commerce activity stream $E_t$. We are, in fact, interested in aggregated volumes of specific EPA types and consider a time frame $T=[t_p, t_{p+h}]$, where the date $t_p$ is the day of prediction and $t_{p+h}$ is the prediction horizon $h=1,3,7$ (1-day, 3-day, 7-day ahead). Therefore, we aim to learn a proxy function $f_v$ that at $t_p$ 
defines a cumulative function:
$$F_V(b_i, T, S, E) = \sum_{t=t_p}^{t=t_{p+h}} f_v(b_i, t, S_t, E_t)$$
integrating, i.e., summing up $f_v$ over $T$.  

\subsection{Multi-class Predictions}
Instead of predicting specific values of $F_V(b_i, S, T, E_t)$ we predict the distribution of $F_V$ in terms of quantile levels that $F_V$ may attain on a given day or a given period of time. Evaluating the performance of quantile predictors enables us to assess whether different social media signals are predictive of  specific EPA levels. We cast that as a multi-class categorization problem using supervised learning where the number of quantiles $q$ determines the number of classes.  Given a number of quantiles $q$ and a training set with values $F_V(b_i, S, T, E_t)$ from the training time period, we determine the data points $F_V(b_i, S, T, E_t)$ within each quantile, i.e.,  
$$Pr(F_V(b_i, S, T, E_t) \leq f_k) = k/q, \  \forall k \ \in [1, q-1]$$ 
and the corresponding ranges of $F_V$ values. The minimum/maximum values determine the thresholds for assigning a class label to an instance of $F_V$.

Training of the multi-class predictor is based on partitioning the training set into distribution quantiles with corresponding class labels and value ranges. We use the quantile ranges to assign class labels to $F_V$ values of the test set. If the test distribution is broader than the one of the training set, $F_V$ values below the minimum or above the maximum are assigned to the lowest/highest quantile, respectively.  

Given a time of prediction $t_p$, we extract features from the social media stream $S$ up to $t_p$ and predict $F_V$ for the prediction horizon $t_{p+h}$. We consider activities on a daily basis and use $t_{p}$ as 23:59:59 on each day.


\subsection{Feature Construction}
Our feature set is based on the three types of SMA signals: Post, Repost and Comment activities. For each signal type we generate features by calculating statistics from SMA streams for the time spans of a pre-specified length. In particular, we consider K-day periods with $K{\in}\{1,3,5,7\}$. Table~\ref{Tab:feature} shows the statistics that we calculate for each signal type, i.e., Post, Repost, Comment, over the specific stream length $K$. The Theil-Sen estimator \cite{akritas1995theil} is a non-parametric trend detector that corresponds to the median over all possible combinations of slopes over $K$ daily measures of the SMA activity. In total, we construct 22 features for each of the 3 SMA types; the total of 66 features that characterize a vendor's SMA.

\begin{table}[h]
\captionsetup{justification=centering}
\caption{Description of statistics used as features for each SMA type over $K$ days before $t_p$. For $K=1$ we consider only $Feat_{sum}$.}\label{Tab:feature}
\centering
\begin{tabular}{lp{6.3cm}}
\hline
Feature& Description\\ [0.5ex] 
\hline
$Feat_{sum}$& Sum of activity volume over K-day period\\
$Feat_{mean}$& Mean activity volume over K-day period\\
$Feat_{max}$& Maximum activity volume over K-day period\\
$Feat_{min}$& Maximum activity volume over K-day period\\
$Feat_{var}$& Variance of activity volume over K-day period\\
$Feat_{stdev}$& Standard deviation of activity volume over K-day period \\
$Feat_{prev}$&Activity volume on previous day\\
$Feat_{theil}$& Theil-Sen estimator\\
\hline
\end{tabular}
\end{table}

\section{Experiments} \label{sec:exp}

We use Logistic Regression and Random Forest \cite{hosmer2013applied,yu2011dual,liaw2002classification} to learn multi-class classification models for 2, 3, and 5 quantiles (2-q, 3-q, 5-q) and vary the prediction horizon by modeling next day (1-day), 3-day and 7-day cumulative volumes of Search, Clickthrough and Orders. We evaluate multi-class prediction results by calculating standard precision, recall and F1 statistics. However, we focus our discussion on the precision since the aim is to assess the effectiveness of SMA features in predicting volume levels of specific EPA types. Thus, identifying \textit{true-positive} instances within a specific quantile is given a priority over avoiding \textit{false-negative} instances. In fact, correct predictions of \textit{high} (top 20\%-30\%) and \textit{low} (bottom 20\%-30\%) quantiles are of particular interest since their detection can improve SMA campaigns and optimize e-commerce operations for high levels of activities that on some days, such as global shopping events, can increase 100-fold. 

\begin{figure}[h]
\captionsetup{justification=centering}
    \centering
    \textbf{}\par\medskip
	\includegraphics[width=8cm,height=8cm,keepaspectratio]{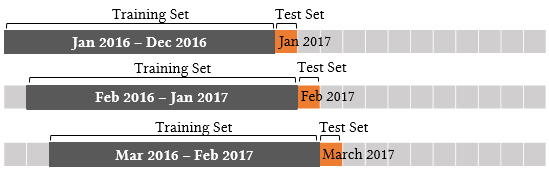}
    \caption{Sliding window of a 12-month training and 1-month test data with a shift of one calendar month, to cover the test period from Jan 2017 until Dec 2017.}\label{Fig:training_testing}
\end{figure}

\begin{figure*}[h]
\captionsetup{justification=centering}
    \centering
    \begin{subfigure}[t]{0.48\textwidth}
        \centering
        \includegraphics[width=\linewidth]{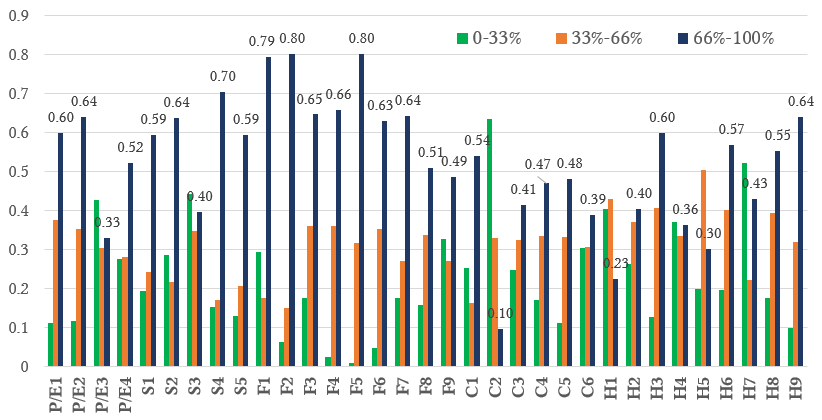} 
        \caption{Precision statistics for 3-q next-day predictions for Orders} \label{fig:1dayOrder3Quantile}
    \end{subfigure}
    \hfill
    \begin{subfigure}[t]{0.48\textwidth}
        \centering
        \includegraphics[width=\linewidth]{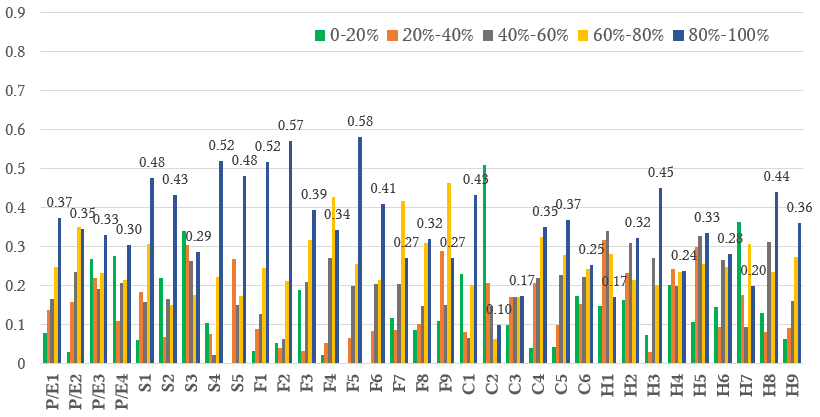} 
        \caption{Precision statistics for 5-q next-day predictions for Orders} \label{fig:1dayOrder5Quantile}
    \end{subfigure}    
    \caption{Precision of Random Forest predictors for 3-q and 5-q categories for the next-day Orders across 33 vendors. The top quantile precisions are higher than random predictions (33\% for 3-q and 20\% for 5-q).}\label{fig:quantileprediction}
\end{figure*}

\begin{table*}[h]
\captionsetup{justification=centering}
\caption{Precision statistics for Random Forest 5-q classifiers for the next-day volumes of each EPA type: Order, Clickthrough and Search. Results are aggregated per five categories. \iffalse\textcolor{red}{import correct statistics}\fi}\label{Tab:5quantile_result_rf}
\centering
\scalebox{0.74}{
\begin{tabular}{ll|P{1cm}|P{1cm}|P{1cm}|P{1cm}|P{1cm}|P{1cm}|P{1cm}|P{1cm}|P{1cm}|P{1cm}|P{1cm}|P{1cm}|P{1cm}|P{1cm}|P{1cm}}
\hline

\multicolumn{2}{c|}{\multirow{2}{*}{Vendor}} & \multicolumn{3}{c|}{0-20\%} & \multicolumn{3}{c|}{20\%-40\%} & \multicolumn{3}{c|}{40\%-60\%} & \multicolumn{3}{c|}{60\%-80\%} & \multicolumn{3}{c}{80\%-100\%}\\\cline{3-17}

& &AVG&MAX&MIN&AVG&MAX&MIN&AVG&MAX&MIN&AVG&MAX&MIN&AVG&MAX&MIN\\
\hline\hline
\multirow{5}{*}{Order} &P\&E&0.159&0.338&0.013&0.136&0.218&0.054&0.198&0.292&0.070&0.298&0.570&0.149&\textbf{0.341}&\textbf{0.393}&\textbf{0.271}\\

&Sports&0.129&0.340&0.000&0.161&0.370&0.045&0.154&0.263&0.021&0.247&0.425&0.149&\textbf{0.456}&\textbf{0.653}&\textbf{0.154}\\
&Food&0.070&0.231&0.000&0.090&0.302&0.000&0.144&0.271&0.030&0.313&0.462&0.181&\textbf{0.410}&\textbf{0.708}&\textbf{0.213}\\
&Clothes&0.199&0.571&0.000&0.129&0.217&0.011&0.186&0.385&0.065&0.225&0.464&0.018&\textbf{0.308}&\textbf{0.528}&\textbf{0.076}\\
&Home&0.167&0.596&0.000&0.157&0.317&0.000&0.263&0.517&0.092&0.222&0.333&0.091&\textbf{0.312}&\textbf{0.481}&\textbf{0.079}\\
\hline\hline

\multirow{5}{*}{Clickthrough} &P\&E&\textbf{0.371}&\textbf{0.548}&\textbf{0.192}&0.200&0.311&0.099&0.147&0.241&0.025&0.159&0.353&0.000&0.265&0.565&0.055\\

&Sports&\textbf{0.462}&\textbf{0.631}&\textbf{0.275}&0.264&0.393&0.094&0.144&0.328&0.038&0.135&0.348&0.024&0.148&0.286&0.037\\
&Food&\textbf{0.351}&\textbf{0.860}&\textbf{0.083}&0.223&0.488&0.056&0.177&0.345&0.030&0.136&0.290&0.000&0.162&0.385&0.036\\
&Clothes&\textbf{0.312}&\textbf{0.571}&\textbf{0.013}&0.165&0.260&0.030&0.199&0.426&0.012&0.145&0.308&0.000&0.192&0.500&0.023\\
&Home&\textbf{0.324}&\textbf{0.645}&\textbf{0.100}&0.225&0.347&0.116&0.188&0.385&0.025&0.149&0.259&0.039&0.174&0.365&0.000\\

\hline\hline

\multirow{5}{*}{Search} &P\&E&\textbf{0.521}&\textbf{0.679}&\textbf{0.336}&0.231&0.353&0.065&0.109&0.293&0.029&0.077&0.176&0.000&0.255&0.481&0.070\\

&Sports&\textbf{0.483}&\textbf{0.763}&\textbf{0.276}&0.248&0.350&0.103&0.130&0.222&0.000&0.081&0.173&0.000&0.126&0.250&0.066\\
&Food&\textbf{0.308}&\textbf{0.784}&\textbf{0.089}&0.193&0.370&0.027&0.183&0.324&0.027&0.175&0.373&0.052&0.185&0.317&0.048\\
&Clothes&\textbf{0.329}&\textbf{0.628}&\textbf{0.194}&0.251&0.353&0.114&0.232&0.313&0.131&0.127&0.238&0.020&0.119&0.288&0.043\\
&Home&\textbf{0.351}&\textbf{0.573}&\textbf{0.179}&0.232&0.323&0.041&0.200&0.322&0.083&0.200&0.311&0.098&0.194&0.316&0.086\\

\hline

\end{tabular}
}
\end{table*}

\subsubsection{Temporal Cross-validation} \label{sssec:crossval}
All our experiments are performed using 12-fold time-series cross-validation, as shown in Figure~\ref{Fig:training_testing}, with data sets specific to each vendor. We train and test our models using a two year JD data set (Section ~\ref{ssec:data}). Our starting training set covers the 12 month period from 1 January 2016 to 31 December 2016. The test set is the following month. In each fold, we slide the 12-month training and 1-month test period by one calendar month. Thus, for each experiment we use one year of historical data and predict EPA in every calendar month in 2017. For each vendor we report the average precision statistics across 12-folds. We collate and average precision statistics across 33 vendors and across vendor categories.

\subsubsection{Activity Predictions} \label{sssec:activity_pred}
For a given quantile scale (2-q, 3-q, 5-q) we predict volumes of activities for the next day $t_{p+1}$ (1-day) and the multi-day cumulative activities over 3-day and 7-day periods. For each prediction type and each individual vendor, we use the corresponding quantile scale determined over the training data. Our experiments thus involve 891 predictions: for each of 33 vendors, 3 EPA types (Search, Clickthrough, Orders) with three quantile scales (2-q, 3-q and 5-q) and 3 time periods (1-day, 3-day, 7-day). For this discussion we select experiments that shed light on: 
\begin{itemize}
\item How successful the predictors are in identifying quantiles for individual EPA types across the vendor sample?
\item How well the predictors perform for the cumulative 3-day and 7-day activities across the vendor sample? 
\item Which features significantly contribute to the performance of predictors for the individual EPA type?
\end{itemize}


\begin{table*}[h]
\captionsetup{justification=centering}
\caption{Three quantile prediction results for five categories of products across 33 vendors based on Random Forest, reporting average precision by categories of 1-day (1D), 3-day (3D) and 7-day (7D) cumulative Orders, Clickthrough and Search for each quantile.}\label{Tab:3quantile_result_rf_137d}
\centering
\scalebox{0.96}{
\begin{tabular}{ll|P{1.3cm}|P{1.3cm}|P{1.3cm}|P{1.3cm}|P{1.3cm}|P{1.3cm}|P{1.3cm}|P{1.3cm}|P{1.3cm}}
\hline

\multicolumn{2}{c|}{\multirow{2}{*}{Vendor}} & \multicolumn{3}{c|}{0-33\%} & \multicolumn{3}{c|}{33\%-66\%} & \multicolumn{3}{c}{66\%-100\%}\\\cline{3-11}

& &1D&3D&7D&1D&3D&7D&1D&3D&7D\\
\hline\hline
\multirow{5}{*}{Order} &P/E&0.232&0.225&0.237&0.327&0.329&0.303&\textbf{0.522}&\textbf{0.536}&\textbf{0.551}\\
&Sports&0.241&0.212&0.172&0.237&0.240&0.298&\textbf{0.584}&\textbf{0.598}&\textbf{0.607}\\
&Food&0.141&0.139&0.133&0.288&0.230&0.251&\textbf{0.663}&\textbf{0.656}&\textbf{0.675}\\
&Clothes&0.287&0.286&0.280&0.297&0.272&0.310&\textbf{0.397}&\textbf{0.431}&\textbf{0.478}\\
&Home&0.261&0.266&0.244&0.375&0.383&0.351&\textbf{0.453}&\textbf{0.467}&\textbf{0.464}\\

&AVG&0.226&0.222&0.209&0.310&0.293&0.302&\textbf{0.528}&\textbf{0.540}&\textbf{0.556}\\\cline{2-11}
\hline\hline

\multirow{5}{*}{Clickthrough} &P/E&\textbf{0.488}&\textbf{0.480}&\textbf{0.471}&0.269&0.303&0.232&0.385&0.362&0.372\\
&Sports&\textbf{0.637}&\textbf{0.620}&\textbf{0.623}&0.226&0.263&0.214&0.232&0.223&0.266\\
&Food&\textbf{0.464}&\textbf{0.471}&\textbf{0.476}&0.282&0.285&0.286&0.263&0.266&0.285\\
&Clothes&\textbf{0.418}&\textbf{0.399}&\textbf{0.423}&0.301&0.311&0.293&0.331&0.328&0.329\\
&Home&\textbf{0.475}&\textbf{0.487}&\textbf{0.484}&0.327&0.307&0.289&0.285&0.278&0.256\\
&AVG&\textbf{0.488}&\textbf{0.486}&\textbf{0.490}&0.288&0.295&0.271&0.291&0.286&0.293\\

\hline\hline

\multirow{5}{*}{Search} &P/E&\textbf{0.597}&\textbf{0.621}&\textbf{0.655}&0.182&0.194&0.150&0.277&0.259&0.289\\
&Sports&\textbf{0.621}&\textbf{0.613}&\textbf{0.616}&0.266&0.248&0.263&0.166&0.157&0.152\\
&Food&\textbf{0.442}&\textbf{0.413}&\textbf{0.440}&0.316&0.319&0.292&0.299&0.291&0.276\\
&Clothes&\textbf{0.455}&\textbf{0.479}&\textbf{0.476}&0.375&0.370&0.355&0.225&0.205&0.172\\
&Home&\textbf{0.453}&\textbf{0.474}&\textbf{0.524}&0.360&0.313&0.298&0.330&0.327&0.326\\
&AVG&\textbf{0.493}&\textbf{0.497}&\textbf{0.522}&0.315&0.301&0.283&0.271&0.261&0.254\\

\hline

\end{tabular}
}
\end{table*} 

\begin{figure*}[h]
\captionsetup{justification=centering}
    \centering
    \begin{subfigure}[t]{0.32\textwidth}
        \centering
        \includegraphics[width=\linewidth]{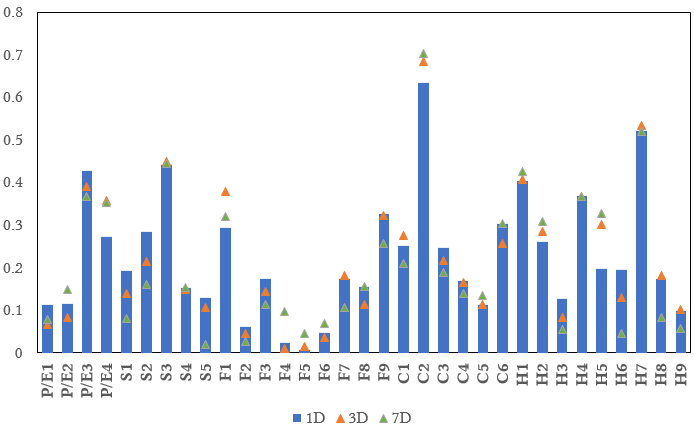} 
        \caption{Precision of bottom quantile predictions for Orders} \label{fig:low_quan}
    \end{subfigure}
    \hfill
    \begin{subfigure}[t]{0.32\textwidth}
        \centering
        \includegraphics[width=\linewidth]{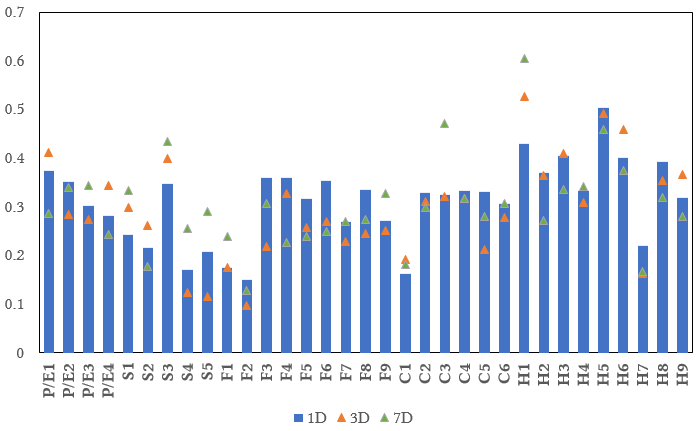} 
        \caption{Precision of middle quantile predictions for Orders} \label{fig:median_quan}
    \end{subfigure}    
    \hfill
    \begin{subfigure}[t]{0.32\textwidth}
        \centering
        \includegraphics[width=\linewidth]{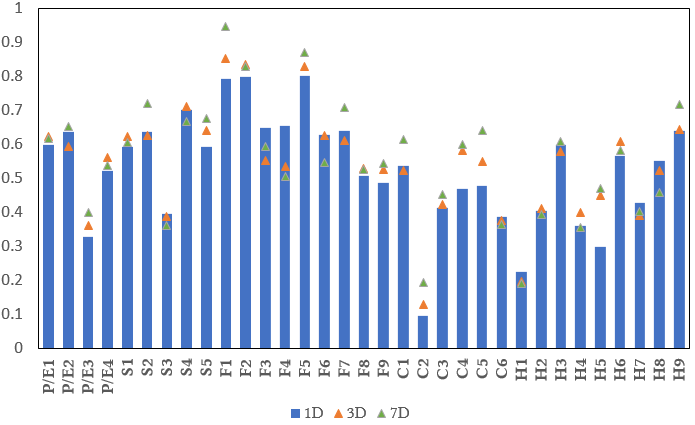} 
        \caption{Precision of top quantile prediction for Orders} \label{fig:high_quan}
    \end{subfigure}
    \caption{Precision of Random Forest predictions into three quantiles for 1-day, 3-day and 7-day cumulative Orders.}\label{fig:3_quan}
\end{figure*}

\subsection{Next-day EPA Predictions for 3 and 5 Quantiles}
Experiments with the next day predictions of EPA for 3-q and 5-q show that the Random Forest (RF) predictors perform higher than random for Orders in the top quantiles (top 33\% and 20\%, respectively). Figure~\ref{fig:quantileprediction} presents RF precision for Order quantiles for individual vendors. Table~\ref{Tab:5quantile_result_rf} summarizes RF precision statistics for all three EPA types and the five vendor categories. We highlight the RF results that are better than random predictions ($>0.20$ for 5-q). We see that, in addition to the top quantile predictions for Orders, the precision statistics are better than random for Search and Clickthrough in the lowest quantile (bottom 33\% for 3-q and bottom 20\% for 5-q). All other quantile predictions for Order, Clickthrough, and Search are on par or lower than random. 

We conducted the same experiments with Logistic Regression (LR) and found that RF outperforms LR predictors. In fact, for our sample of product vendors, the t-test performed for LR and RF predictions yield $p<0.05$ for all the quantile labels (not just the top quantiles).

\subsection{Predictions of Multi-day Cumulative EPA}
Considering a selective success of SMA based predictors for quantiles of the next-day EPA, we train predictors for cumulative EPA, i.e., EPA volumes over 3 and 7 days. We expect that multi-day cumulative statistics are less volatile and therefore the quantile levels may be more stable and predictable over time. 

Our experiments show that RF predictors for 3-day and 7-day cumulative EPA are consistently performing at the similar order of magnitude for a given quantile level. That is illustrated in Table~\ref{Tab:3quantile_result_rf_137d}, showing the precision statistics of RF predictors for 1-day, 3-day and 7-day cumulative volumes of Order, Clickthrough and Search for 3 quantiles (3-q). Similar observations are made for 5 quantiles and for the LR classifier. 

We conclude that our high performing predictors for cumulative multi-day EPA volumes can be flexibly applied to different scenarios that benefit from observing cumulative EPA across multiple days. This includes the precision of top quantile volumes for Orders.  Figure \ref{fig:3_quan} shows 3-q predictions of 1-day, 3-day and 7-day cumulative Orders for individual vendors.

\subsection{Feature Significance}
Both Random Forest and Logistic Regression allow us to assess the significance of individual feature types in terms of their contributions to the prediction decision. For illustration, we present the analysis of features for the RF classifier trained for 3-q predictions of the next-day EPA volumes. For each classifier we calculate the average relative rank of a feature. Table~\ref{Tab:featureimportance} shows the top 10 contributing features of 1-day EPA predictors for Orders, Clickthrough and Search, respectively, across all the vendors and 3 quantiles. 

We observe that the relative feature importance is, to a degree, in agreement with the observations from the correlation analysis in Section~\ref{sssec:activity_pred}. Search levels are predicted by Comment and Repost activities considered over different lengths of time: 1, 3, 5, and 7 days. Clickthrough activities seem to be aligned with SMA over 7 and 3 day periods, primarily with consumer comments. Orders, however, are clearly related to Comment volumes over a longer periods, i.e., 5 and 7 days. Overall, we recognize the importance of features generated from Comment activities, as they contribute to the predictors of all EPA types.

\begin{table}[h]
\captionsetup{justification=centering}
\caption{Ten top-ranked features of the 3-q Random Forest predictors for the next-day EPA.}\label{Tab:featureimportance}
\centering
\scalebox{0.9}{
\begin{tabular}{l|l|l|l}
\hline
Rank & Order& Clickthrough&Search\\ 
\hline\hline
1&$7DComment_{min}$&$7DComment_{min}$&$7DComment_{min}$\\
2&$7DComment_{sum}$&$3DComment_{mean}$&$5DRepost_{min}$\\
3&$7DComment_{mean}$&$3DComment_{sum}$&PreviousDComment\\
4&$7DComment_{max}$&PreviousDComment&$7DComment_{std}$\\
5&$5DComment_{mean}$&$3DComment_{max}$&$3DComment_{mean}$\\
6&$5DComment_{max}$&$7DComment_{sum}$&$7DRepost_{std}$\\
7&$5DComment_{sum}$&$7DComment_{mean}$&$5DRepost_{sum}$\\
8&$7DComment_{std}$&$7DPost_{sum}$&$3DComment_{sum}$\\
9&$7DComment_{var}$&$3DComment_{min}$&$7DComment_{var}$\\
10&$3DComment_{mean}$&$5DComment_{var}$&$5DComment_{var}$\\

\hline
\end{tabular}
}
\end{table}

\section{Concluding Remarks}
This paper presents a detailed empirical study of the relationship between SMA of product vendors and EPA of consumers interested in the vendors' products. The study is the first to characterize the correlation of specific SMA engagements, i.e., posts, reposts and comments, and EPA types that correspond to specific stages in the consumers' purchase decisions, i.e., search for brands and products, clickthrough product information, and product orders. Our analyses uncovered low-to-moderate correlations between the volumes of SMA-EPA pairs, suggesting that predicting daily volumes of EPA based on SMA volumes alone is not well supported. However, moderate correlations of Post-Search, Post-Clickthrough and Comment-Order across vendors, suggest that one may be able to train predictors of EPA distributions and their changes rather that the precise daily volumes. We thus introduce a new approach of characterizing SMA-EPA relationship in terms of EPA quantiles. We formulate EPA predictions as a  multi-class categorization problem into 2, 3 and 5 quantiles. Our Random Forest classifiers outperform both the random predictors and the Logistic Regression classifiers for top quantiles of Order and bottom quantiles of Search and Clickthrough. Our study provides unique insights into the varied correlations of SMA-EPA pairs and a mixed success of SMA-based predictors in determining EPA levels. A general view that social media engagements undoubtedly drive consumer traction on e-commerce platforms is not substantiated by the Sina Weibo and JD case study and requires further analysis.  
In our future work we will expand explorations of this issue with a broader set of product categories and SMA analyses that consider properties of the content exchanged through the SMA.

\section{Acknowledgements}

We wish to express our gratitude to JD.com for providing access to invaluable data and Bin Xu and Hao Dong for their assistance. We also acknowledge the financial support from the International Doctoral Innovation Centre, Ningbo Education Bureau, Ningbo Science and Technology Bureau, and the University of Nottingham. This work was also supported by the UK Engineering and Physical Sciences Research Council [grant number EP/L015463/1].

\bibliographystyle{ieeetr}
\bibliography{references}

\end{document}